\documentclass[journal]{IEEEtran}

\usepackage{cite}
\usepackage{amsmath,amssymb,amsfonts}
\usepackage{algorithmic}
\usepackage{graphicx}
\usepackage{textcomp}
\usepackage{booktabs}
\usepackage{multirow}
\usepackage{array}
\usepackage{url}

\def\BibTeX{{\rm B\kern-.05em{\sc i\kern-.025em b}\kern-.08em
  T\kern-.1667em\lower.7ex\hbox{E}\kern-.125emX}}

\begin{document}
\bstctlcite{BSTcontrol}

\title{Decision-Aligned Value Learning for Selective Adoption of EEG Fusion in Human-Robot Collaboration}

\author{Xinjia Yu\textsuperscript{1},
Jing Yang\textsuperscript{1},
Yang Zhou\textsuperscript{1},
Tielin Shi\textsuperscript{1},
and Tao Cheng\textsuperscript{2,*}%
\thanks{\textsuperscript{1}School of Mechanical Science and Engineering, Huazhong University of Science and Technology, 1037 Luoyu Road, Wuhan 430074, China.}%
\thanks{\textsuperscript{2}School of Urban Transport and Logistics, Shenzhen Technology University, 3002 Lantian Road, Shenzhen 518118, China.}%
\thanks{\textsuperscript{*}Corresponding author: Tao Cheng (e-mail: chengtao@sztu.edu.cn).}}

\maketitle

\begin{abstract}
Electroencephalography (EEG) captures neural activity associated with human perception and decision-making. It provides human-robot collaboration (HRC) systems with measurable evidence for event recognition, situation assessment, and action preparation. External sensing cannot easily provide such evidence. However, the quality and complementarity of this evidence vary across samples, and existing EEG fusion methods do not explicitly determine when EEG improves a prediction. We therefore framed selective EEG use as a sample-wise incremental-value estimation problem and proposed Value Estimation and Constraint-Aware Triage for Adoption (VECTA). VECTA separates the generation of an EEG-informed Fusion prediction from the decision about whether it should replace the existing Base prediction. The framework uses system gain to quantify the loss reduction from Base to Fusion and matching gain to assess whether benefits arise from sample-matched EEG. A signed-value learner estimates system gain from decision-time evidence and prioritizes high-value Fusion outputs under budget constraints. Cross-subject evaluations on three HRC datasets showed that matched Fusion outperformed shuffled controls. Estimated gains tracked realized gains, with Spearman correlations of 0.347–0.612 and positive-gain areas under the receiver operating characteristic curve (AUROCs) of 0.708–0.842. Under budget constraints, VECTA reduced Brier loss and harmful Fusion adoption and achieved the highest mean normalized AUBC across three datasets among the selective-decision methods evaluated. Validation on three external multi-modal datasets showed that VECTA could be applied directly as a post-hoc decision layer to task-specific predictors without retraining. These results show that selective Fusion adoption can translate the uneven utility of EEG into system-level gains.
\end{abstract}


\section{Introduction}
\label{sec:introduction}

\IEEEPARstart{I}{n} industrial human-robot collaboration (HRC), safe and efficient coordination requires robots to recognize workers' ongoing actions and understand a broader range of information about their human partners \cite{elgohr2026trust,li2023proactive}. Relevant human information includes whether the worker is cognitively ready and able to continue safely as task demands change \cite{yao2024task}, understands the current task context, stage, and interaction turn \cite{ehrlich2023human,ma2025human}, has perceived task-relevant events that may not be fully captured by robot sensing or predefined rules \cite{john2024prediction,buerkle2021feasibility,li2024brain}, and is preparing a particular action \cite{khanna2026early}. Access to such information complements established sensing pathways and provides a more complete and reliable basis for understanding the human partner and supporting collaborative decisions \cite{kalinowska2023embodied,wang2024multimodal}. These factors concern internal human states, task awareness, cognitive appraisal, and emerging intentions. Existing HRC perception systems primarily rely on vision, motion, force, and system-state signals to characterize the external environment and capture externally observable information about the human partner, such as actions, postures, movement trajectories, and physical interactions with the robot \cite{saleem2025review,angleraud2024sensor,jung2024arbitrary}. However, these interactions are largely external manifestations of ongoing perceptual, cognitive, and physiological processes. Conventional sensing focuses on these manifestations, but much of the information about the underlying internal states and evolving processes remains insufficiently captured \cite{hoffman2024inferring,elgohr2026trust}.

Electroencephalography (EEG) has been widely studied in HRC because it captures neural activity with high temporal resolution \cite{zhang2025mind}. Compared to peripheral physiological signals, EEG provides more direct access to perceptual and cognitive processes and to some related internal information that may not be fully expressed through speech, actions, or explicit commands \cite{wu2023affective}. These properties make EEG a promising sensing modality. Despite this potential, EEG is not uniformly reliable. It is affected by a low signal-to-noise ratio, artifacts, intersubject variability, temporal nonstationarity, and task-dependent changes in neural responses \cite{yilmaz2026eeg}. Its contribution also varies across samples. For some samples, EEG improves the existing prediction. For others, it adds little value or leads to a worse decision than the established system would have made without EEG \cite{hao2026step}. Such sample-level heterogeneity calls for the selective adoption of EEG-informed Fusion. 

Existing EEG-assisted HRC studies have demonstrated the task-level utility of neural information through task-relevant decoding and multi-modal prediction \cite{yu2027dsmt,john2024prediction,xu2025neutrl,khanna2026early,xue2026multimodal}. These findings show that EEG can provide useful task-level information. The next step is to translate this utility into sample-wise decisions. This raises two related questions. The first is whether the incremental value of EEG over an established non-EEG predictor can be estimated for each sample using information available at decision time. The second is whether these estimates can be used to determine which Fusion outputs to adopt under a prescribed budget. Addressing these questions can reduce the effects of sample-level unreliability and focus EEG use on the decisions it is most likely to improve. This downstream allocation problem is closely related to established paradigms for selective decision-making, such as selective prediction, learning to defer, and algorithmic triage \cite{geifman2019selectivenet,mozannar2020consistent,okati2021differentiable}. Selective prediction adds an abstention mechanism to a predictor to manage the trade-off between predictive risk and coverage \cite{traub2024overcoming}. Learning to defer determines whether each instance should be handled by the model or delegated to an expert, depending on which is expected to perform better \cite{narasimhan2022posthoc}. Algorithmic triage distributes a limited decision budget across automated and expert decisions to optimize overall system performance \cite{okati2021differentiable,zhang2026coverage}. However, the positive incremental value of EEG cannot be fully characterized by the confidence of either prediction alone \cite{jitkrittum2023confidence}. This value is the reduction in predictive loss when the EEG-informed Fusion prediction replaces the Base prediction. Despite their principled foundation for sample-wise allocation, the standard formulations of these paradigms rarely account explicitly for this nested relationship. The incremental value also depends on the unknown ground-truth outcome and is therefore unavailable when the adoption decision must be made. This challenge calls for a dedicated decision model to guide the selective adoption of EEG-informed Fusion under such uncertainty.

To address this problem, we proposed Value Estimation and Constraint-Aware Triage for Adoption (VECTA), a decision-aligned framework for the selective adoption of EEG-informed Fusion predictions. VECTA learns to estimate the sample-wise system gain of adopting Fusion at decision time. Under a prescribed adoption budget, it ranks samples by their estimated system gain and adopts the Fusion predictions for the highest-ranked samples. The main contributions of this work are summarized as follows:

\begin{itemize}
\item We formulated the question of when EEG-informed Fusion should influence the final decision as a sample-wise incremental-value estimation problem. We introduced system gain to quantify the benefit of Fusion over Base and guide adoption, and matching gain to independently audit the contribution of sample-matched EEG evidence.

\item We developed VECTA to separate the generation of EEG-informed predictions from the decision to adopt them. Using the information available at decision time, the signed-value learner estimated how much the prediction loss for each sample could decrease or increase if Base were replaced with Fusion. The resulting gain estimates guided adoption under a given budget.

\item We demonstrated the broader applicability of VECTA by evaluating it on three primary HRC datasets and validating it downstream as a post-hoc decision layer over frozen multi-modal predictors on three external datasets. The downstream results show that VECTA yields measurable system-level gains when EEG provides identifiable complementary evidence and conservatively retains Base when positive EEG value is sparse. These findings support the use of VECTA as a value-guided decision layer in existing HRC prediction systems.

\end{itemize}

\section{Methodology}
\label{sec:method}

\subsection{Problem Formulation and Value Definition}
\label{sec:Problem Formulation}
\begin{figure}[!t]
\centering
\includegraphics[width=\columnwidth]{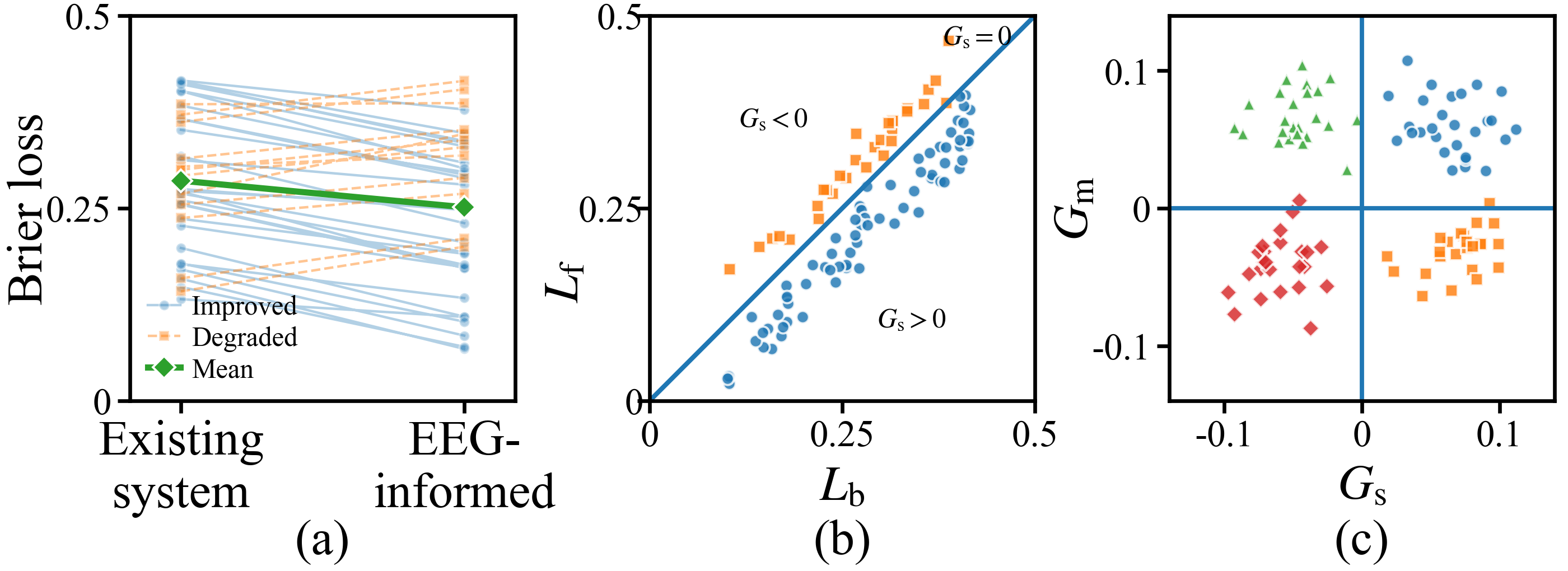}
\caption{Sample-wise characterization of EEG value. (a) Comparison of aggregate Brier loss with and without EEG information. (b) Geometry of system gain in the space of reference and EEG-informed losses. (c) Joint distribution of system gain \(G_{\mathrm{s}}\) and matching gain \(G_{\mathrm{m}}\).}
\label{fig:value_definition}
\end{figure}
\begin{figure*}[!t]
\centering
\includegraphics[width=\textwidth]{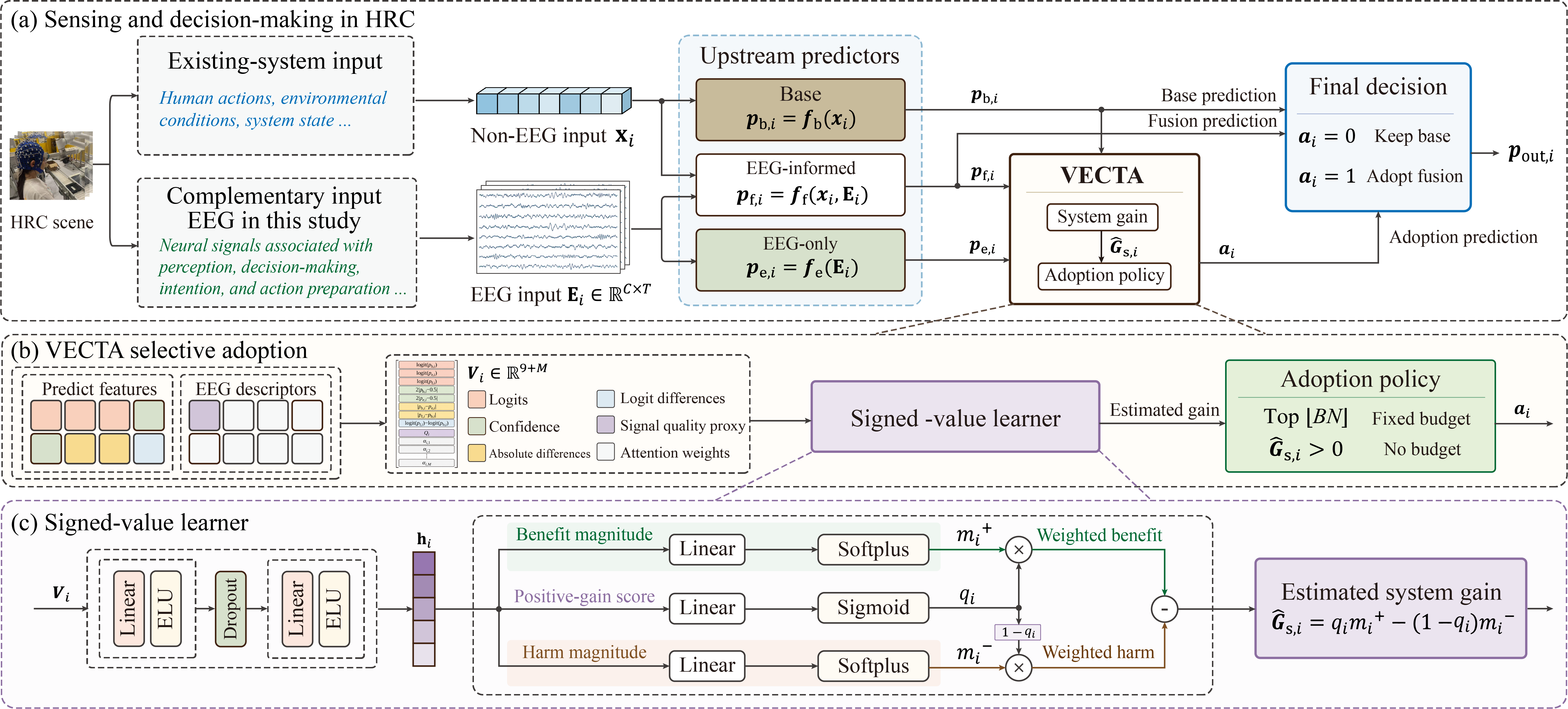}
\caption{VECTA Framework for Selective Adoption of EEG-Informed Fusion Predictions in HRC.}
\label{fig:vecta}
\end{figure*}
The contribution of EEG is often evaluated using aggregate predictive performance, typically by comparing system configurations with and without EEG information \cite{khanna2026early,xue2026multimodal}. This comparison indicates whether EEG improves performance on average, but it does not show how its effect varies across individual samples. Fig.~\ref{fig:value_definition}(a) illustrates this discrepancy. Although the mean Brier loss decreases after EEG is introduced, the loss increases for some individual samples. Thus, an average performance improvement does not imply that the EEG-informed prediction should be adopted for every sample.

This motivates a shift from aggregate EEG performance to its sample-wise incremental value. For sample \(i\), let \(p_{\mathrm{b},i}\) denote the prediction produced by the existing perception system and \(p_{\mathrm{f},i}\) denote the corresponding EEG-informed prediction. Their Brier losses \cite{tyralis2024review} are
\(L_{\mathrm{b},i}=\ell_{\mathrm{Br}}(p_{\mathrm{b},i},y_i)\) and
\(L_{\mathrm{f},i}=\ell_{\mathrm{Br}}(p_{\mathrm{f},i},y_i)\), where
\(\ell_{\mathrm{Br}}(p,y)=(p-y)^2\). At the sample level, an accuracy difference records only whether the predicted class changes from correct to incorrect or vice versa. It cannot distinguish changes in probabilistic prediction quality when both predictions yield the same class. An explicit quantitative measure is therefore needed to characterize the sample-wise value of EEG. We define the \emph{system gain} as the sample-wise reduction in prediction loss achieved by replacing the existing system prediction with its EEG-informed counterpart:
\begin{equation}
G_{\mathrm{s},i}=L_{\mathrm{b},i}-L_{\mathrm{f},i}.
\label{eq:system_gain}
\end{equation}
Accordingly, \(G_{\mathrm{s},i}>0\) means that adopting the EEG-informed prediction improves the current system decision, whereas \(G_{\mathrm{s},i}<0\) indicates a detrimental effect. The magnitude \(|G_{\mathrm{s},i}|\) quantifies the size of this change. Fig.~\ref{fig:value_definition}(b) shows this relationship. The diagonal \(L_{\mathrm{b}}=L_{\mathrm{f}}\) corresponds to zero gain, while samples below and above the diagonal have positive and negative system gains, respectively. Unlike a binary indicator of benefit or harm, \(G_{\mathrm{s}}\) captures both the direction and magnitude of the EEG-induced change. This finer-grained measure is important when only a limited number of EEG-informed decisions can be adopted, because samples must be prioritized according to how much benefit EEG is expected to provide.

System improvement alone does not fully characterize the role of EEG. A lower Fusion loss does not necessarily mean that the improvement depends on the EEG input paired with the current sample. To construct a sample-specific matching reference, we randomly permute the EEG inputs across samples while retaining the non-EEG input for sample \(i\). Let \(p_{\mathrm{f},i}^{(\pi_k)}\) denote the Fusion prediction obtained under the \(k\)th random permutation. The matching gain is defined as
\begin{equation}
G_{\mathrm{m},i}
=
\frac{1}{K}\sum_{k=1}^{K}
\left(p_{\mathrm{f},i}^{(\pi_k)}-y_i\right)^2
-
\left(p_{\mathrm{f},i}-y_i\right)^2.
\label{eq:matching_gain}
\end{equation}
A positive \(G_{\mathrm{m},i}\) indicates that correctly paired EEG produces a lower Brier loss than mismatched EEG. This provides evidence that the Fusion benefit depends on sample-specific neural information. A non-positive value indicates that correct pairing provides no measurable advantage over the shuffled reference. Unlike \(G_{\mathrm{s},i}\), which measures whether Fusion improves on Base, \(G_{\mathrm{m},i}\) is an attribution measure and does not directly determine whether Fusion should be adopted.

Fig.~\ref{fig:value_definition}(c) shows four possible situations. First, an EEG-informed decision may provide both a system benefit and a benefit associated with correctly matched EEG. Second, system performance may improve even when the matching benefit is weak. Third, correctly matched EEG may contain useful sample-specific information but fail to improve the final decision. Fourth, neither type of benefit may be observed. Together, \(G_{\mathrm{s}}\) and \(G_{\mathrm{m}}\) extend the characterization of EEG value from a one-dimensional measure of system improvement to a two-dimensional description. The two dimensions address distinct questions: \emph{whether} EEG improves the decision and \emph{whether} that improvement is attributable to correctly matched, sample-specific neural information.

\subsection{VECTA: Sample-Wise System Gain Estimation}
\label{sec:System Gain}

The system gain \(G_{\mathrm{s},i}\), defined in Sec.~\ref{sec:Problem Formulation}, depends on the ground-truth label \(y_i\) and is therefore unavailable at decision time. VECTA addresses this limitation by estimating \(G_{\mathrm{s},i}\) from observable information for the current sample. Fig.~\ref{fig:vecta}(b) shows the estimation workflow. The existing-system, EEG-only, and EEG-informed predictors first produce their outputs. These outputs are combined with descriptors of EEG quality and temporal contribution to form a value feature vector. A signed-value learner then maps this vector to an estimated system gain \(\widehat{G}_{\mathrm{s},i}\). This estimate represents the expected reduction in prediction loss if the EEG-informed prediction is adopted.

Let \(p_{\mathrm{b},i}\), \(p_{\mathrm{e},i}\), and \(p_{\mathrm{f},i}\) denote the probabilities produced by the existing-system, EEG-only, and EEG-informed predictors, respectively. VECTA constructs the following value feature vector:
\begin{equation}
\begin{aligned}
\mathbf{v}_i=\big[&
\operatorname{logit}(p_{\mathrm{b},i}),
\operatorname{logit}(p_{\mathrm{e},i}),
\operatorname{logit}(p_{\mathrm{f},i}),\\
&
2|p_{\mathrm{b},i}-0.5|,
2|p_{\mathrm{e},i}-0.5|,
|p_{\mathrm{b},i}-p_{\mathrm{e},i}|,
|p_{\mathrm{f},i}-p_{\mathrm{b},i}|,\\
&
\operatorname{logit}(p_{\mathrm{f},i})
-\operatorname{logit}(p_{\mathrm{b},i}),
Q_i,
\alpha_{i,1},\ldots,\alpha_{i,M}
\big]^{\top}
\in\mathbb{R}^{9+M}.
\end{aligned}
\label{eq:value_features}
\end{equation}
Here, \(\operatorname{logit}(p)=\log[p/(1-p)]\). For numerical stability, \(p\) is clipped to \([10^{-5},1-10^{-5}]\). The first three elements encode the direction and strength of the three predictor outputs. The next two measure the confidence of the existing-system and EEG-only predictions based on their distances from the decision threshold. The differences \(|p_{\mathrm{b},i}-p_{\mathrm{e},i}|\) and \(|p_{\mathrm{f},i}-p_{\mathrm{b},i}|\) characterize the consistency between the available predictors and the extent to which EEG changes the original system prediction. The corresponding logit difference preserves the direction and scale of this EEG-induced correction. \(Q_i\) is an EEG-quality descriptor. We use the RMS amplitude as a lightweight proxy for signal quality, following established EEG quality-assessment practice \cite{guggenberger2026realtime}. The terms \(\alpha_{i,1},\ldots,\alpha_{i,M}\) are the normalized attention weights assigned to the \(M\) EEG temporal windows and satisfy \(\sum_{m=1}^{M}\alpha_{i,m}=1\).

Fig.~\ref{fig:vecta}(c) shows the signed-value learner. A two-layer multilayer perceptron with ELU activations maps the value feature vector \(\mathbf{v}_i\) to a latent representation \(\mathbf{h}_i\). Dropout is applied after the first layer. Three linear output heads then use \(\mathbf{h}_i\) to to produce a positive-gain score of a positive system gain and the conditional magnitudes of positive and non-positive outcomes:
\begin{equation}
q_i
=
\sigma\left(
\mathbf{w}_{q}^{\top}\mathbf{h}_i+b_q
\right).
\label{eq:gain_sign}
\end{equation}
\begin{equation}
\begin{aligned}
m_i^{+}
&=
\operatorname{softplus}\left(
\mathbf{w}_{+}^{\top}\mathbf{h}_i+b_{+}
\right),\\
m_i^{-}
&=
\operatorname{softplus}\left(
\mathbf{w}_{-}^{\top}\mathbf{h}_i+b_{-}
\right).
\end{aligned}
\label{eq:gain_magnitudes}
\end{equation}
Here, \(q_i\) is a bounded gain score that adopting the EEG-informed prediction produces a positive system gain. Conditional on the gain direction, \(m_i^{+}\) estimates the expected improvement when EEG is beneficial, whereas \(m_i^{-}\) estimates the expected deterioration when EEG provides no positive gain.

These outputs capture complementary aspects of sample-wise EEG value. The positive-gain score alone cannot distinguish a small improvement from a substantial one. It also cannot account for the severity of a potentially harmful adoption. In contrast, directly regressing the signed gain can average positive and negative effects toward zero and underrepresent their different consequences. VECTA therefore uses separate outputs for the positive-gain score and the conditional magnitudes of benefit and harm. These components are combined as
\begin{equation}
\widehat{G}_{\mathrm{s},i}
=
q_i m_i^{+}
-
(1-q_i)m_i^{-}.
\label{eq:expected_system_gain}
\end{equation}

The resulting score has a direct decision interpretation. It weighs the potential benefit of adopting Fusion against its potential harm. Thus, two samples with similar positive-gain scores can receive different values when their potential improvements or adverse effects differ. This formulation aligns the value learner with budget-constrained adoption. Under a limited budget, samples must be ranked by their expected system-level contribution.

\subsection{Budget-Constrained EEG-Fusion Adoption}
\label{sec:budgeted_adoption}

Given a candidate set \(\mathcal{I}=\{1,\ldots,N\}\) and a fixed Fusion adoption budget \(B\in[0,1]\), VECTA ranks the samples in descending order of \(\widehat{G}_{\mathrm{s},i}\) \cite{okati2021differentiable}. Fixing the adoption rate separates selection quality from adoption volume and allows controlled comparisons across operating points. The budget can also represent an operational policy that limits the proportion of final decisions influenced by an auxiliary physiological modality.

For a given budget \(B\), let \(K_B=\lfloor BN\rfloor\) be the number of samples for which Fusion can determine the final output. Let \(\mathcal{S}_B\subseteq\mathcal{I}\) contain the indices of the \(K_B\) largest estimated system gains. The corresponding adoption indicator is \(a_i(B)=\mathbb{I}(i\in\mathcal{S}_B)\). Under the fixed-cardinality constraint \(\sum_i a_i(B)=K_B\), this Top-\(K_B\) rule maximizes the cumulative estimated system gain \(\sum_i a_i(B)\widehat{G}_{\mathrm{s},i}\). Thus, under the fixed-budget evaluation protocol, selection depends on the rank of \(\widehat{G}_{\mathrm{s},i}\) across samples, not only on whether the estimated gain is positive. In deployment, \(B\) can instead be treated as an upper bound. Samples with \(\widehat{G}_{\mathrm{s},i}\leq 0\) are then rejected even if the available quota is not fully used. When no adoption budget is imposed, Fusion is adopted only for samples with \(\widehat{G}_{\mathrm{s},i}>0\).

The final system prediction uses the Fusion output for selected samples and retains the Base output for all remaining samples:
\begin{equation}
p_{\mathrm{out},i}=a_i(B)p_{\mathrm{f},i}+\left[1-a_i(B)\right]p_{\mathrm{b},i}.
\label{eq:budgeted_output}
\end{equation}
Accordingly, \(B=0\) yields the Base prediction for every sample, whereas \(B=1\) corresponds to Always Fusion. At intermediate budgets, EEG-informed predictions are adopted according to their estimated system-level contributions.

To examine whether matching-specific value can be inferred from information available at decision time, an independent copy of the signed-value learner is trained using \(G_{\mathrm{m},i}\) as its target. This produces the auxiliary estimate \(\widehat{G}_{\mathrm{m},i}\). The auxiliary branch uses the same value-feature construction and signed-value architecture as the main system-gain estimator but is optimized independently. Its estimates are used to assess the predictability of matching gain. The realized \(G_{\mathrm{m},i}\) values for held-out samples are used to audit whether the samples prioritized by VECTA benefit from correctly paired EEG. Neither \(\widehat{G}_{\mathrm{m},i}\) nor the realized \(G_{\mathrm{m},i}\) enters the adoption policy. All routing decisions are based only on \(\widehat{G}_{\mathrm{s},i}\).

\subsection{Decision-Aligned Value Learning}
\label{subsec:decision_aligned_learning}

In each outer leave-one-subject-out (LOSO) split, one subject is held out for final testing, and the remaining subjects are divided into \(F\) inner subject folds \cite{gil2026across}. For each inner fold, the upstream predictor is trained on the other \(F-1\) folds and applied to the held-out subjects to generate Base, EEG-only, matched-Fusion, and shuffled-Fusion predictions. The \(F\) sets of out-of-fold (OOF) predictions are concatenated. They are then divided by subject into fitting and validation subsets for VECTA optimization and early stopping, respectively. This procedure ensures that each prediction record is generated by an upstream predictor that was not trained on the corresponding subject. After the OOF predictions are generated, the upstream predictor is retrained on all outer-training subjects and applied to the outer test subject. Thus, the upstream predictor is fitted \(F+1\) times in each outer split.

Let \(z_i=\mathbb{I}(G_{\mathrm{s},i}>0)\). The sign output \(q_i\) is trained using class-weighted binary cross-entropy. The positive and negative magnitude outputs, \(m_i^{+}\) and \(m_i^{-}\), are trained using Smooth-L1 loss on samples with \(z_i=1\) and \(z_i=0\), respectively. Their targets are \(G_{\mathrm{s},i}\) and \(-G_{\mathrm{s},i}\). Since budgeted adoption also depends on the relative ordering of samples, VECTA minimizes \cite{burges2005learning}
\begin{equation}
\mathcal{L}_{\mathrm{rank}}=\frac{1}{|\mathcal{R}|}\sum_{(i,j)\in\mathcal{R}}\operatorname{softplus}\left[-\frac{s_{ij}\left(\widehat{G}_{\mathrm{s},i}-\widehat{G}_{\mathrm{s},j}\right)}{\tau_{\mathrm{r}}}\right], \label{eq:ranking_loss}
\end{equation}
where \(\mathcal{R}\) contains sampled pairs satisfying \(|G_{\mathrm{s},i}-G_{\mathrm{s},j}|>\delta\), \(s_{ij}=\operatorname{sign}(G_{\mathrm{s},i}-G_{\mathrm{s},j})\), and \(\tau_{\mathrm{r}}\) is a temperature parameter. The resulting estimation objective is \(\mathcal{L}_{\mathrm{est}}=\lambda_{\mathrm{sign}}\mathcal{L}_{\mathrm{sign}}+\lambda_{\mathrm{mag}}\mathcal{L}_{\mathrm{mag}}+\lambda_{\mathrm{rank}}\mathcal{L}_{\mathrm{rank}}\). The three terms respectively model whether EEG is beneficial, estimate the magnitude of its effect, and learn which samples should receive higher adoption priority.

Because the hard Top-\(K_B\) rule is non-differentiable, VECTA uses a differentiable decision-level proxy to incorporate the downstream prediction loss into training:
\begin{equation}
\widetilde{a}_i=q_i\,\sigma\left(\frac{\widehat{G}_{\mathrm{s},i}}{\tau_{\mathrm{g}}}\right),\ \widetilde{p}_i=p_{\mathrm{b},i}+\widetilde{a}_i\left(p_{\mathrm{f},i}-p_{\mathrm{b},i}\right), \label{eq:soft_adoption}
\end{equation} 
where \(\tau_{\mathrm{g}}>0\) controls the sharpness of the soft transition, \(\widetilde{a}_i\) is the soft-adoption weight, and \(\widetilde{p}_i\) is the corresponding system prediction. For the \(N\) samples in the fitting subset, let \(\ell_{\mathrm{pol},i}=\ell_{\mathrm{Br}}(\widetilde{p}_i,y_i)\) and \(\overline{a}=N^{-1}\sum_i\widetilde{a}_i\). Let \(u\) index the training subjects, and let \(\mathcal{I}_u\) contain the samples from subject \(u\). The mean policy loss for subject \(u\) is \(\overline{\ell}_u=|\mathcal{I}_u|^{-1}\sum_{i\in\mathcal{I}_u}\ell_{\mathrm{pol},i}\). The decision-level terms are \(\mathcal{L}_{\mathrm{policy}}=N^{-1}\sum_i\ell_{\mathrm{pol},i}\), \(\mathcal{L}_{\mathrm{budget}}=[\max(0,\overline{a}-B_0)]^2\), and \(\mathcal{L}_{\mathrm{group}}=\operatorname{Std}_u(\overline{\ell}_u)\). These terms respectively optimize the soft system output, penalize adoption rates above the training-time upper bound \(B_0\), and reduce variation across subjects. Here, \(B_0\) is a fixed upper bound used only to regularize soft adoption during training. In contrast, \(B\) specifies the hard Top-\(K_B\) adoption rate during evaluation. The same trained value model is evaluated across all values of \(B\). The decision-level terms are combined with \(\mathcal{L}_{\mathrm{est}}\) using the weights \(\lambda_{\mathrm{policy}}\), \(\lambda_{\mathrm{budget}}\), and \(\lambda_{\mathrm{group}}\). The soft-adoption mechanism is used only during training. Inference follows the hard adoption rule.

\section{Experimental Results and Discussion}
\label{sec:results and discussion}

\subsection{Experimental Setup}
\label{sec:experimental_setup}
We evaluated VECTA on three public EEG datasets involving human--robot interaction. ErrP-CoAdaptation \cite{ehrlich2018coadaptation} examined participants' perception of errors in targets selected by a robot. It contained 4,790 samples from 16 participants, of which 2,399 were evaluation samples with an error rate of 46.23\%. ErrP-HRI \cite{ehrlich2019feasibility} focused on distinguishing correct from erroneous system outcomes following observed robot actions. Of its 11,435 samples from 12 participants, 9,162 were evaluation samples with a system-error rate of 35.32\%. HRC-NextActor \cite{ehrlich2023human} was designed to predict the next actor (the human or the robot) during collaborative task planning. It contained 21,815 samples from 11 participants, of which 17,452 were evaluation samples with a 51.75\% Human-class rate. The Supplementary Material provides details of the dataset-specific preprocessing, feature construction, and upstream predictions. Together, these datasets covered human perceptions of machine decisions, task performance, and upcoming collaborative actions. This coverage enabled VECTA to select reliable EEG evidence to help the system better understand human cognitive states.

We used outer LOSO evaluation, with 16, 12, and 11 folds for ErrP-CoAdaptation, ErrP-HRI, and HRC-NextActor, respectively. For inner subject-level cross-fitting, we set \(F = 3\). We assigned 20\% of the subjects in each outer-training set to the VECTA validation set for early stopping. Each outer fold was repeated using seeds 42, 52, and 62, resulting in 117 runs. We estimated all normalization statistics exclusively from the outer-training data and did not use any held-out samples for personalization.

VECTA used a two-layer MLP with a hidden dimension of 48; both were trained with AdamW at a learning rate of \(10^{-3}\) and subject-disjoint early stopping. The matching gain was estimated from eight within-subject and within-stage EEG shifts. Exact-quota budgets were \(B \in \{20\%,40\%,60\%,80\%,100\%\}\), with 40\% as the primary setting. VECTA was compared to five established selective-decision methods: SelectiveNet \cite{geifman2019selectivenet}, L2D \cite{mozannar2020consistent}, LUT \cite{okati2021differentiable}, post-hoc L2D \cite{narasimhan2022posthoc}, and CL2DC \cite{zhang2026coverage}. Supplementary Section SI-C provides details of their implementations and experimental settings. After averaging the results across seeds for each subject, we performed paired comparisons using two-sided Wilcoxon tests with Holm correction and reported rank-biserial effect sizes. The 95\% confidence intervals were estimated using 10,000 subject-level bootstrap resamples.

\subsection{Predictive Utility of EEG and Matched Fusion}
\label{sec:predictive_utility}
\begin{table}[t]
\centering
\caption{Predictive Performance Across Three Datasets}
\label{tab:predictive_utility}
\footnotesize
\setlength{\tabcolsep}{2.5pt}
\renewcommand{\arraystretch}{1.08}
\begin{tabular*}{\columnwidth}{@{\extracolsep{\fill}}llcc@{}}
\hline
\textbf{Dataset} & \textbf{Predictor} &
\textbf{AUROC (\%)} $\uparrow$ &
\textbf{Brier} $\downarrow$ \\
\hline
\multirow{4}{*}{\shortstack[l]{ErrP-\\CoAdaptation}}
& Base & 71.79 $\pm$ 14.76 & 0.2024 $\pm$ 0.0613 \\
& EEG-only & 77.40 $\pm$ 11.67 & 0.2043 $\pm$ 0.0460 \\
& Shuffled Fusion & 67.01 $\pm$ 12.74 & 0.2373 $\pm$ 0.0624 \\
& Always Fusion & \textbf{83.02} $\pm$ \textbf{11.03}
& \textbf{0.1672} $\pm$ \textbf{0.0618} \\
\hline
\multirow{4}{*}{ErrP-HRI}
& Base & 49.33 $\pm$ 2.23 & 0.2556 $\pm$ 0.0041 \\
& EEG-only & 75.76 $\pm$ 8.59 & 0.1966 $\pm$ 0.0330 \\
& Shuffled Fusion & 50.20 $\pm$ 0.93 & 0.2997 $\pm$ 0.0211 \\
& Always Fusion & \textbf{76.90} $\pm$ \textbf{8.32}
& \textbf{0.1928} $\pm$ \textbf{0.0286} \\
\hline
\multirow{4}{*}{\shortstack[l]{HRC-\\NextActor}}
& Base & 85.85 $\pm$ 1.26 & 0.1422 $\pm$ 0.0070 \\
& EEG-only & 61.42 $\pm$ 3.47 & 0.2454 $\pm$ 0.0079 \\
& Shuffled Fusion & 84.41 $\pm$ 1.35 & 0.1522 $\pm$ 0.0081 \\
& Always Fusion & \textbf{86.67} $\pm$ \textbf{1.78}
& \textbf{0.1416} $\pm$ \textbf{0.0098} \\
\hline
\end{tabular*}
\vspace{2pt}
\end{table}
Table~\ref{tab:predictive_utility} assesses the predictive utility of EEG and the effect of correct EEG--sample pairing on fusion performance. The main text reports AUROC and Brier score to characterize discrimination and probabilistic prediction accuracy, respectively. Supplementary Table SIII provides the complete results for accuracy, Macro-F1, and ECE. EEG-only achieved AUROCs of 77.40\%, 75.76\%, and 61.42\% on the three datasets. These results support the presence of task-relevant information in EEG.

Always Fusion significantly outperformed Shuffled Fusion in AUROC and Brier score across all three datasets. The complete five-metric comparisons also favored correct EEG--sample pairing. Compared to Base, Always Fusion significantly improved both metrics on ErrP-CoAdaptation and ErrP-HRI. However, the differences on HRC-NextActor were nonsignificant. Together, the three datasets represented settings with clear average fusion benefits and settings with limited gains beyond a strong base predictor.

\subsection{Sample-Wise Value Estimation and Routing Performance}
\label{sec:routing_performance}
\subsubsection{Sample-Wise Gain Heterogeneity}
\begin{figure*}[!t]
\centering
\includegraphics[width=\textwidth]{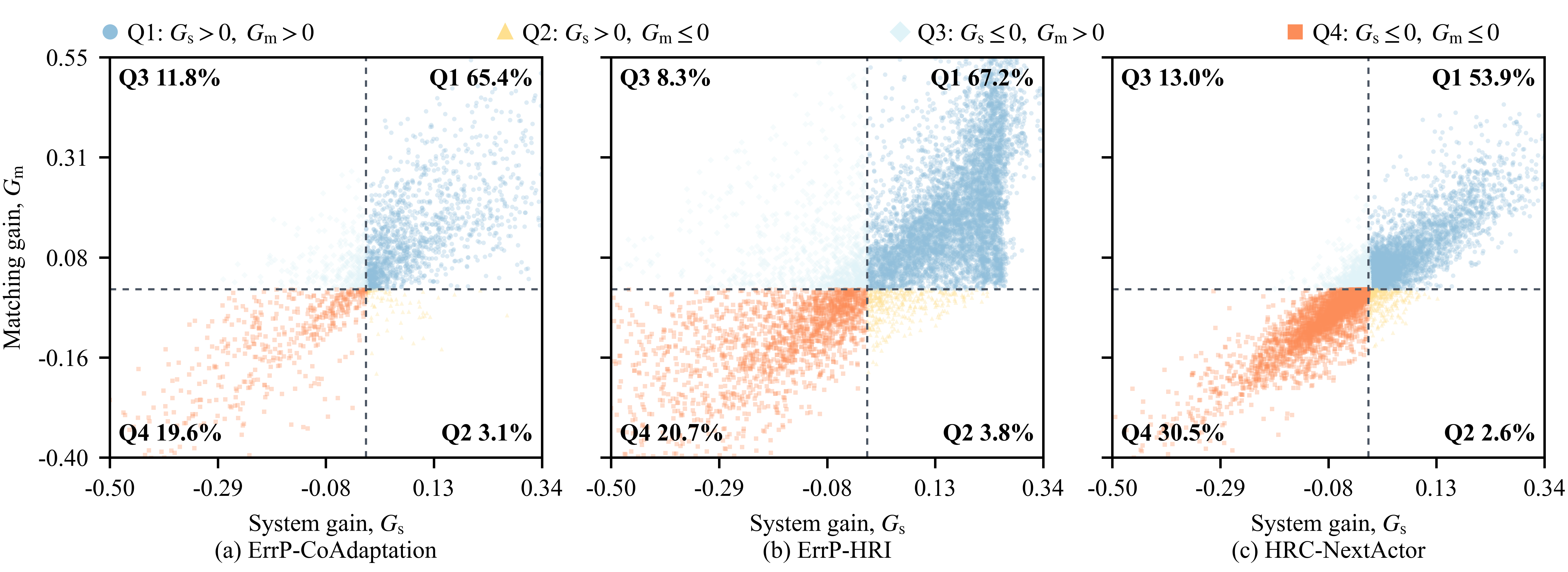}
\caption{Distributions of the system and matching gains across the three datasets. Each point shows the gains for one query sample, with the values averaged across three random seeds. Quadrant proportions were calculated separately for each subject and then averaged across subjects.}
\label{fig:dual_true_gains}
\end{figure*}
Before evaluating routing performance, we examined how fusion value varied across samples and datasets. \(G_{\mathrm{s}}\) quantifies system improvement but cannot attribute that improvement to the EEG paired with the current sample. We therefore examined \(G_{\mathrm{m}}\) alongside it solely as a matching-specific audit. Fig.~\ref{fig:dual_true_gains} shows the joint distributions of the two realized gains.

ErrP-CoAdaptation and ErrP-HRI exhibited clear average benefits from fusion. Their mean \((G_{\mathrm{s}}, G_{\mathrm{m}})\) values were \((0.0352, 0.0701)\) and \((0.0627, 0.1068)\), respectively. For ErrP-CoAdaptation, the 95\% confidence intervals for \(G_{\mathrm{s}}\) and \(G_{\mathrm{m}}\) were \([0.0155, 0.0564]\) and \([0.0493, 0.0921]\), respectively. For ErrP-HRI, they were \([0.0477, 0.0795]\) and \([0.0850, 0.1322]\), respectively. In the two datasets, Q1 accounted for 65.40\% and 67.21\% of the samples. Thus, most samples benefited at the system level and from correct EEG pairing, with both gains above zero. However, Q4 still contained 19.64\% and 20.68\% of the samples, showing that the positive average gains concealed a number of harmful or non-beneficial fusion cases.

HRC-NextActor showed a different pattern. Its mean system gain was only 0.0006, with a 95\% confidence interval of \([-0.0027, 0.0035]\). Its mean matching gain, however, remained positive at 0.0106, with a confidence interval of \([0.0084, 0.0128]\). Although Q1 still accounted for 53.92\% of the samples, Q4 increased to 30.55\%. Beneficial fusion cases were therefore present, but harmful cases largely offset the improvements they produced. This explains why Always Fusion provided little average improvement over the strong Base predictor.

The two gains were closely related but not interchangeable. Their median within-subject Spearman correlations were 0.810, 0.881, and 0.907 for ErrP-CoAdaptation, ErrP-HRI, and HRC-NextActor, respectively. Samples with larger system gains therefore generally also benefited more from correct EEG pairing. Nevertheless, the gains differed in sign for 14.96\%, 12.11\%, and 15.54\% of the samples. Q2 was uncommon, accounting for 3.13\%, 3.83\%, and 2.57\%. Thus, few samples improved over Base without a positive matching advantage. Q3 accounted for 11.84\%, 8.28\%, and 12.97\%, showing that positive matching gains failed to translate into system improvement for only a relatively small proportion of samples. Overall, beneficial and non-beneficial fusion cases coexisted across all three datasets. This coexistence confirms the existence and importance of the problem addressed by VECTA and establishes the need for sample-wise selection.

\subsubsection{Sample-Wise Gain Prediction and Ranking}
\begin{table}[t]
\centering
\caption{Sample-Wise Gain-Prediction Performance}
\label{tab:value_prediction}
\footnotesize
\setlength{\tabcolsep}{2pt}
\renewcommand{\arraystretch}{1.08}
\begin{tabular*}{\columnwidth}{@{\extracolsep{\fill}}llccc@{}}
\hline
\textbf{Dataset} & \textbf{Target} &
\textbf{Spearman} $\uparrow$ &
\textbf{MAE} $\downarrow$ &
\textbf{AUROC} $\uparrow$ \\
\hline
\multirow{2}{*}{\shortstack[l]{ErrP-\\CoAdaptation}}
& $G_{\mathrm{s}}$ & 0.347 $\pm$ 0.221 & 0.116 $\pm$ 0.033 & 0.747 $\pm$ 0.113 \\
& $G_{\mathrm{m}}$ & 0.147 $\pm$ 0.128 & 0.133 $\pm$ 0.037 & 0.630 $\pm$ 0.079 \\
\hline
\multirow{2}{*}{ErrP-HRI}
& $G_{\mathrm{s}}$ & 0.612 $\pm$ 0.117 & 0.161 $\pm$ 0.027 & 0.708 $\pm$ 0.055 \\
& $G_{\mathrm{m}}$ & 0.376 $\pm$ 0.169 & 0.178 $\pm$ 0.027 & 0.662 $\pm$ 0.029 \\
\hline
\multirow{2}{*}{\shortstack[l]{HRC-\\NextActor}}
& $G_{\mathrm{s}}$ & 0.445 $\pm$ 0.038 & 0.038 $\pm$ 0.008 & 0.842 $\pm$ 0.015 \\
& $G_{\mathrm{m}}$ & 0.350 $\pm$ 0.049 & 0.044 $\pm$ 0.009 & 0.780 $\pm$ 0.022 \\
\hline
\end{tabular*}
\vspace{2pt}

\end{table}
\begin{figure}[!t]
\centering
\includegraphics[width=\columnwidth]{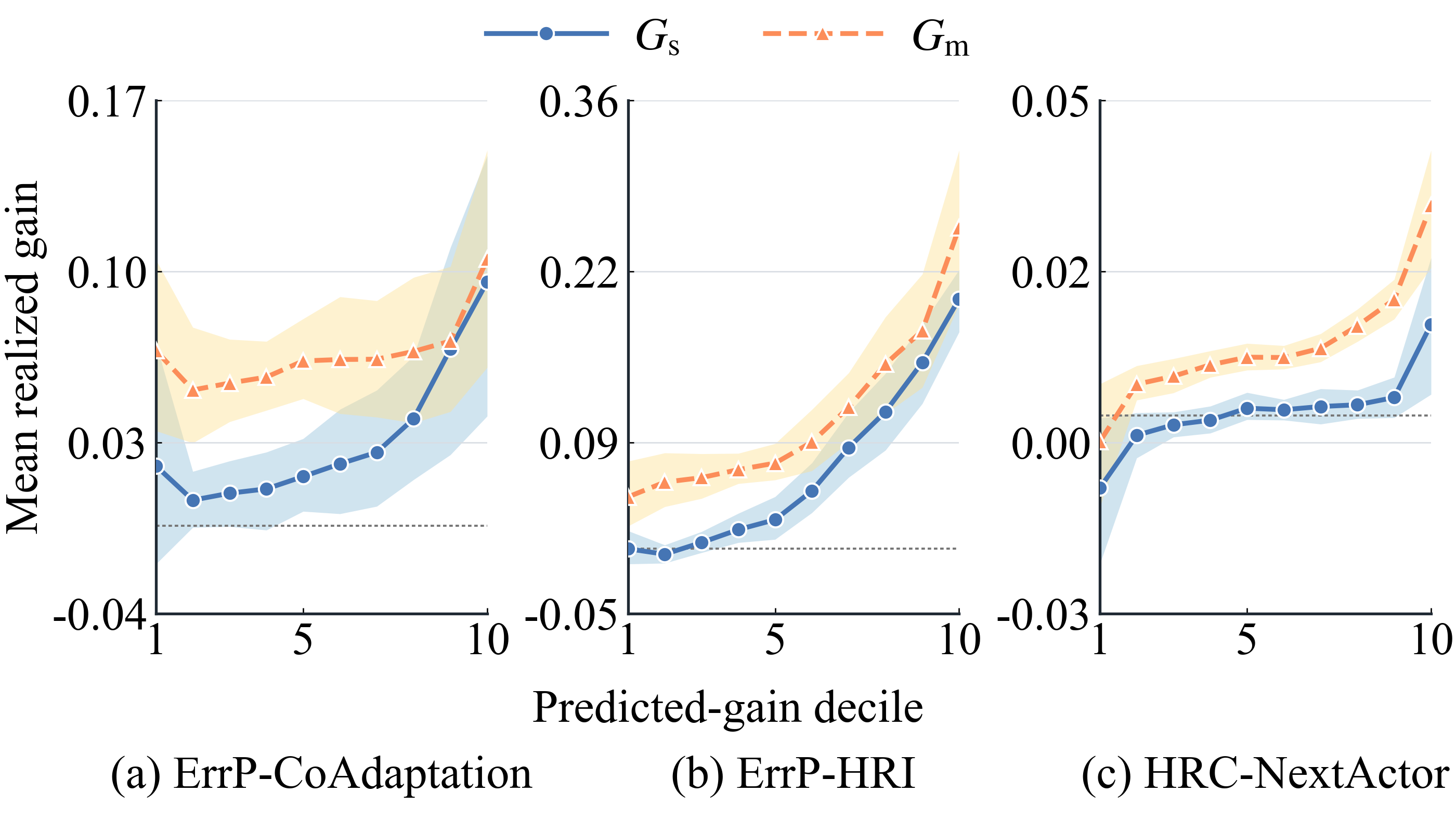}
\caption{Realized system and matching gains across within-subject deciles formed by ranking samples according to \(\widehat{G}_{\mathrm{s}}\) and \(\widehat{G}_{\mathrm{m}}\), respectively. The points show participant-level mean gains, and the shaded regions show 95\% confidence intervals obtained by bootstrapping participants. The horizontal dashed lines indicate zero gain.}
\label{fig:value_deciles}
\end{figure}
We next evaluated the ability of VECTA  to recover this value structure by identifying beneficial samples and correctly ranking samples by gain magnitude. VECTA achieved significant positive correlations between estimated and realized system gains in all three datasets (all Holm-adjusted \(p \leq 0.003\)). ErrP-CoAdaptation showed a moderate mean gain of \(0.0352\), modest ranking accuracy (\(\rho = 0.347\)), and useful positive-gain discrimination (AUROC \(= 0.747\)). ErrP-HRI showed the largest mean gain (\(0.0627\)), the strongest ranking performance (\(\rho = 0.612\)), and moderate discrimination (AUROC \(= 0.708\)). HRC-NextActor showed a negligible mean gain (\(0.0006\)), moderate ranking performance (\(\rho = 0.445\)), and the strongest discrimination (AUROC \(= 0.842\)).

Fig.~\ref{fig:value_deciles} shows the realized gains across predicted-gain deciles. Increasing curves indicate correct ordering, and narrower confidence bands indicate greater stability across participants. The ErrP-CoAdaptation curves were relatively flat across the lower and middle deciles and increased mainly in the highest deciles. Their broad confidence bands indicate substantial variation across participants. ErrP-HRI showed the clearest monotonic increases. This pattern indicates stable overall ordering. However, the widening upper confidence bands indicate variable high-gain magnitudes. HRC-NextActor showed smaller but increasing gains. Its narrow middle bands indicate stable central separation, but its wider tails indicate greater uncertainty at the extremes. Overall, VECTA recovered meaningful gain ordering across all three datasets. The results also showed dataset-dependent differences in discrimination and cross-participant stability across gain levels.

\subsubsection{Budgeted Routing Performance}
\begin{table}[t]
\centering
\caption{Fusion-Benefit Concentration at the 40\% Adoption Budget}
\label{tab:routing_performance}
\footnotesize
\setlength{\tabcolsep}{2pt}
\renewcommand{\arraystretch}{1.08}
\begin{tabular*}{\columnwidth}{@{\extracolsep{\fill}}lccc@{}}
\hline
\textbf{Dataset} & \begin{tabular}[c]{@{}c@{}}\textbf{Always Fusion}\\\textbf{BLR}\end{tabular} & \begin{tabular}[c]{@{}c@{}}\textbf{VECTA BLR}\\\textbf{at 40\%}\end{tabular} & \begin{tabular}[c]{@{}c@{}}\textbf{Retained}\\\textbf{Benefit}\end{tabular} \\
\hline
\begin{tabular}[c]{@{}l@{}}ErrP-\\CoAdaptation\end{tabular} & 0.0352 & 0.0244 & $\approx 69.3\%$ \\
ErrP-HRI & 0.0627 & 0.0543 & $\approx 86.7\%$ \\
\begin{tabular}[c]{@{}l@{}}HRC-\\NextActor\end{tabular} & 0.0006 & 0.0020 & $>100\%$ \\
\hline
\end{tabular*}
\end{table}
\begin{figure}[t]
\centering
\includegraphics[width=\columnwidth]{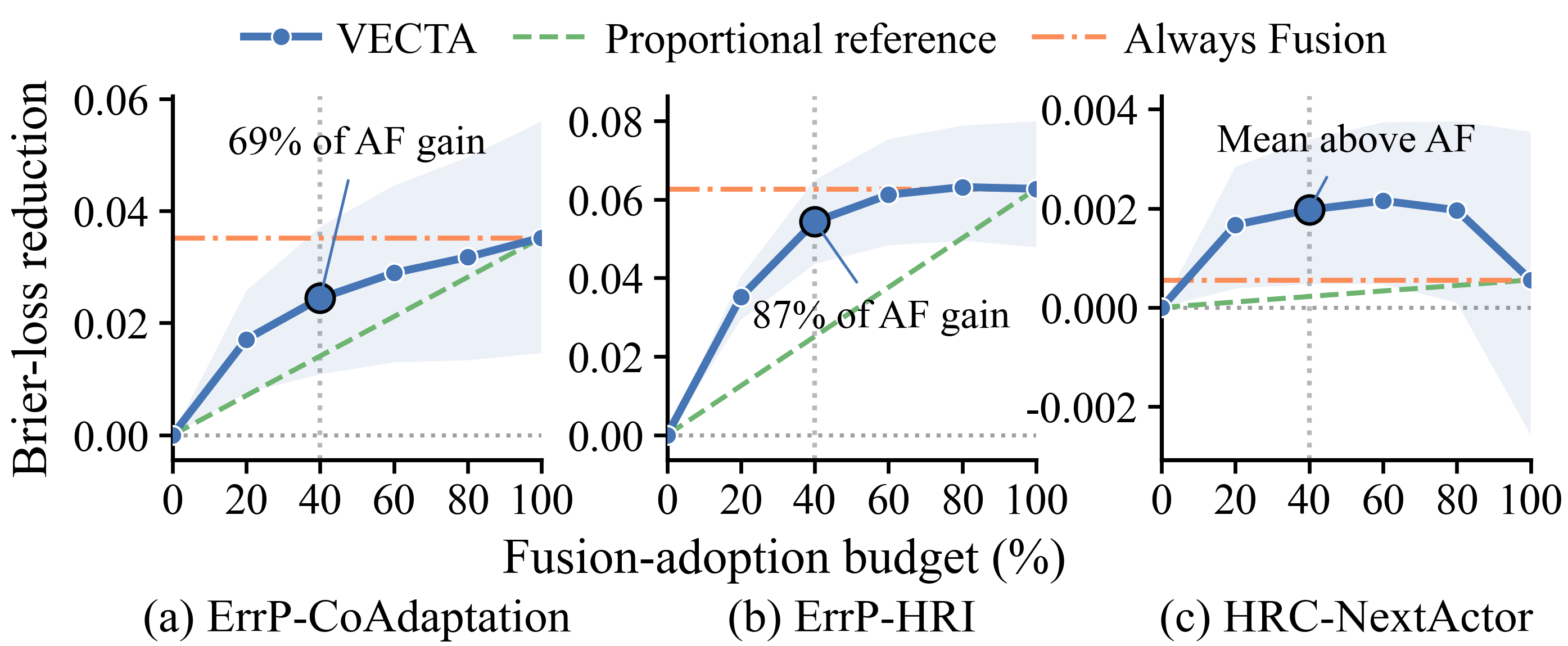}
\caption{Brier-loss reduction achieved by VECTA across Fusion-adoption budgets. The curves show participant-level means with 95\% bootstrap confidence intervals. The dashed lines show the proportional and Always-Fusion references. The y-axis scales differ.}
\label{fig:routing_budgets}
\end{figure}
We next evaluated how effectively VECTA concentrated the benefits of Fusion in a limited adoption quota. For each budget \(B\), \(B\,\mathrm{BLR}_{\mathrm{AF}}\) served as the subject-wise proportional reference. At the primary \(40\%\) budget, VECTA retained \(69.3\%\) and \(86.7\%\) of the mean Always-Fusion gain for ErrP-CoAdaptation and ErrP-HRI, respectively. For HRC-NextActor, its mean BLR was \(0.001975\), compared with \(0.000557\) under Always Fusion (Table~\ref{tab:routing_performance}). VECTA exceeded the proportional reference in all three datasets (all \(p_{\mathrm{Holm}} \leq 0.0076\), \(r_{\mathrm{rb}} = 0.735\text{--}1.000\)). Thus, the selected \(40\%\) of samples captured a disproportionately large share of the full-adoption gain.

Across budgets, Fig.~\ref{fig:routing_budgets} shows that VECTA accumulated gain faster than the proportional reference at low and intermediate adoption rates. For ErrP-HRI and HRC-NextActor, however, its performance flattened or declined at higher budgets. This pattern indicates that expanding the quota progressively admitted samples with smaller or negative gains. These results show that VECTA prioritizes higher-value Fusion outputs and limits gain dilution from less useful EEG evidence.

\subsubsection{Audit of Negative System and Matching Gains}
Because aggregate BLR can conceal harmful samples in the selected set, we audited Harmful-S (\(G_{\mathrm{s}} < 0\)) and Harmful-M (\(G_{\mathrm{m}} < 0\)). Because VECTA uses only \(\widehat{G}_{\mathrm{s}}\) for routing, Harmful-S measures the direct effect of its routing policy, whereas Harmful-M provides a secondary audit of the contribution from matched EEG. At the \(40\%\) budget, VECTA reduced Harmful-S compared to uniform random selection in all three datasets (\(p_{\mathrm{Holm}} \leq 0.015\)). It also reduced Harmful-M in ErrP-HRI and HRC-NextActor (\(p_{\mathrm{Holm}} \leq 0.002\)). The selected sets in these two datasets also contained more Q1 samples, for which both gains were positive (Supplementary Table~SV). These patterns generally persisted at intermediate budgets (Supplementary Fig.~S1). Thus, routing based on the estimated system gain consistently limited harmful Fusion adoption.

\subsection{Comparison with Alternative Selectors}
\label{sec:alternative selectors}
\begin{table}[!t]
\centering
\caption{Mean Normalized AUBC of Alternative Selectors (\%)}
\label{tab:alternative_methods}
\footnotesize
\renewcommand{\arraystretch}{1.12}
\begin{tabular*}{\columnwidth}{@{\extracolsep{\fill}}lccc@{}}
\hline
\textbf{Method} & \textbf{ErrP-CoAdaptation} & \textbf{ErrP-HRI} & \textbf{HRC-NextActor} \\
\hline
SelectiveNet & 2.231 & 2.859 & 0.035 \\
L2D & 2.205 & 5.659 & 0.174 \\
LUT & 2.357 & 5.562 & 0.141 \\
Post-hoc L2D & 2.265 & 5.544 & 0.075 \\
CL2DC & 2.470 & 4.888 & 0.035 \\
\textbf{VECTA} & \textbf{2.781} & \textbf{5.691} & \textbf{0.180} \\
\hline
\end{tabular*}
\end{table}
We compared VECTA with adapted versions of five established selective-decision methods under identical upstream predictions, subject splits, routing inputs, and adoption budgets. Table~\ref{tab:alternative_methods} reports the mean normalized AUBC over the \(20\%\)--\(100\%\) budget range. Supplementary Table~SIV reports ACC and BLR at each budget. VECTA achieved the highest mean normalized AUBC in all three datasets, with values of 2.781, 5.691, and 0.180 for ErrP-CoAdaptation, ErrP-HRI, and HRC-NextActor, respectively. In paired subject-level comparisons of normalized AUBC, VECTA significantly outperformed SelectiveNet and CL2DC on ErrP-HRI (\(p_{\mathrm{Holm}} = 0.0024\) and \(0.0039\), respectively). It also significantly outperformed SelectiveNet, LUT, Post-hoc L2D, and CL2DC on HRC-NextActor (\(p_{\mathrm{Holm}} = 0.0146\), \(0.0371\), \(0.0049\), and \(0.0049\), respectively). All other comparisons, including every comparison on ErrP-CoAdaptation, were nonsignificant. The budget-specific results further showed that VECTA was particularly competitive at the \(60\%\) and \(80\%\) budgets. It achieved the highest BLR at both budgets on ErrP-CoAdaptation and HRC-NextActor and at the \(80\%\) budget on ErrP-HRI. VECTA also achieved the highest BLR at the \(40\%\) budget on ErrP-HRI.

The cross-dataset pattern further clarifies when VECTA is most useful. Its clearest advantage on HRC-NextActor supports direct sample-wise gain estimation when EEG effects vary widely and harmful Fusion decisions can offset beneficial ones. On ErrP-HRI, the close performance of L2D suggests that the routing features already allowed a general deferral method to identify many samples for which Fusion was preferable to Base, leaving less room for VECTA to improve. On ErrP-CoAdaptation, the absence of significant differences despite VECTA’s highest mean AUBC indicates competitive but less consistent performance, possibly reflecting greater between-participant variation. Because a single VECTA ranking was used across all budgets, its strong performance at the \(60\%\) and \(80\%\) budgets suggests that the learned ordering remained effective beyond the \(40\%\) training budget, with the clearest support at moderate-to-high budgets. Overall, these comparisons support directly learning the expected loss reduction from replacing Base with Fusion, particularly when EEG effects are uneven across samples.

\subsection{Ablation}
\label{sec:ablation}
\begin{table}[t]
\centering
\caption{Ablation of VECTA Value-Learning Objectives}
\label{tab:ablation}
\footnotesize
\renewcommand{\arraystretch}{1.06}
\resizebox{\columnwidth}{!}{%
\begin{tabular}{@{}llccc@{}}
\hline
\textbf{Dataset} & \textbf{Variant} & \textbf{Spearman} $\uparrow$ & \textbf{BLR} $\uparrow$ & \textbf{Harmful-S (\%)} $\downarrow$ \\
\hline
\multirow{4}{*}{\shortstack[l]{ErrP-\\CoAdapt.}} & Regression & 0.168$\pm$0.131 & 0.0241$\pm$0.0288 & 31.4$\pm$13.4 \\
& Sign+Mag. & 0.054$\pm$0.196 & 0.0177$\pm$0.0247 & \textbf{23.0$\pm$13.7} \\
& w/o Rank & 0.203$\pm$0.115 & 0.0235$\pm$0.0281 & 27.2$\pm$11.5 \\
& \textbf{Full} & \textbf{0.347$\pm$0.221} & \textbf{0.0244$\pm$0.0279} & 23.9$\pm$13.5 \\
\hline
\multirow{4}{*}{ErrP-HRI} & Regression & 0.550$\pm$0.088 & 0.0525$\pm$0.0177 & 14.3$\pm$6.5 \\
& Sign+Mag. & 0.354$\pm$0.128 & 0.0468$\pm$0.0183 & 17.0$\pm$7.4 \\
& w/o Rank & 0.544$\pm$0.081 & 0.0531$\pm$0.0185 & 14.2$\pm$6.8 \\
& \textbf{Full} & \textbf{0.612$\pm$0.117} & \textbf{0.0543$\pm$0.0197} & \textbf{13.7$\pm$7.2} \\
\hline
\multirow{4}{*}{\shortstack[l]{HRC-\\NextActor}} & Regression & 0.166$\pm$0.075 & 0.0007$\pm$0.0016 & 30.5$\pm$5.4 \\
& Sign+Mag. & 0.372$\pm$0.064 & 0.0013$\pm$0.0016 & \textbf{13.7$\pm$2.3} \\
& w/o Rank & 0.416$\pm$0.044 & \textbf{0.0020$\pm$0.0022} & 16.3$\pm$2.5 \\
& \textbf{Full} & \textbf{0.445$\pm$0.038} & \textbf{0.0020$\pm$0.0026} & 16.3$\pm$2.6 \\
\hline
\end{tabular}%
}
\end{table}
Table~\ref{tab:ablation} compares four variants of the VECTA value learner. The direct-regression variant (Regression) uses one scalar output and a regression loss to predict \(G_{\mathrm{s}}\) directly and test the sufficientIt of direct gain regression. The sign-and-magnitude-only variant (Sign+Mag.) retains the three outputs \(q_i\), \(m_i^{+}\), and \(m_i^{-}\) but uses only the sign and magnitude losses to train them. It tests the sign-and-magnitude decomposition without ranking or decision-level supervision. The variant without the ranking loss (w/o Rank) retains the sign, magnitude, policy, budget, and group losses but removes \(\mathcal{L}_{\mathrm{rank}}\). It tests whether explicit ranking supervision improves sample ordering. Full achieved the highest mean \(G_{\mathrm{s}}\) correlation in all three datasets and the highest or tied-highest mean BLR. Sign+Mag. produced the lowest Harmful-S rate on ErrP-CoAdaptation and HRC-NextActor. This result is consistent with its direct sign supervision, which emphasizes separating positive from negative gains. However, using the sign and magnitude objectives alone resulted in weaker gain ordering and smaller system-level benefits from budgeted routing.

Removing \(\mathcal{L}_{\mathrm{rank}}\) consistently reduced the \(G_{\mathrm{s}}\) correlation. The largest decrease occurred on ErrP-CoAdaptation. In contrast, BLR changed only marginally. This difference arises because Spearman correlation evaluates the ordering of all samples. BLR at the \(40\%\) budget, however, changes only when ranking corrections move samples into or out of the selected top \(40\%\). The ranking loss therefore improves global value ordering even when its effect at one routing cutoff is modest.

\subsection{Downstream Validation with Existing Predictors}
\label{sec:downstream validation}
\begin{table}[t]
\centering
\caption{Downstream Performance Under Unconstrained Routing}
\label{tab:external_evaluation}
\footnotesize
\setlength{\tabcolsep}{4pt}
\renewcommand{\arraystretch}{1.10}

\begin{tabular*}{\columnwidth}{@{\extracolsep{\fill}}lccc@{}}
\hline
\textbf{Dataset} & \textbf{Selected samples} & \textbf{BLR} $\uparrow$ & \textbf{$\Delta$ACC (pp)} $\uparrow$ \\
\hline
RITHM & 405/2100 (19.3\%) & 0.00917 (3.62\%) & +4.81 \\
Handover & 18/880 (2.0\%) & 0.00022 (0.10\%) & 0.00 \\
EgoBrain & 73/137 (53.3\%) & 0.00510 (11.67\%) & +2.92 \\
\hline
\end{tabular*}
\end{table}
\begin{figure}[t]
\centering
\includegraphics[width=\columnwidth]{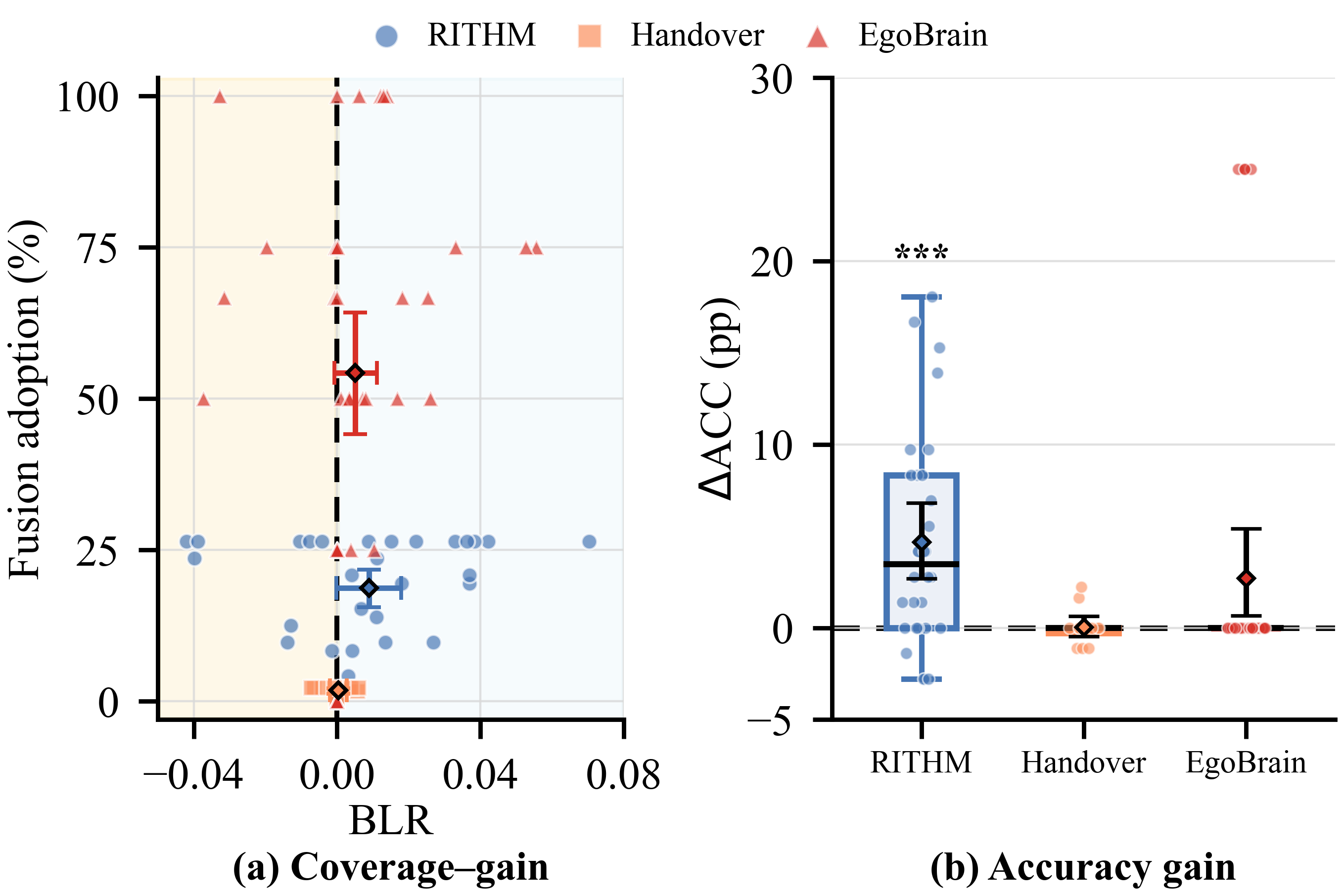}
\caption{Participant-level downstream performance under unconstrained routing. (a) Fusion-adoption coverage versus BLR. (b) Change in accuracy relative to Base. }
\label{fig:downstream_performance}
\end{figure}
We evaluated VECTA as a post-hoc decision layer for task-specific multi-modal predictors on three external multi-modal datasets, with the upstream predictors kept frozen. RITHM--Cognitive Shifting combines EEG and peripheral physiology to monitor cognitive effort during robot-induced task switching \cite{rihet2026robotic}. Human-to-Robot Handover combines EEG, gaze, and hand motion for early recognition of handover intention \cite{khanna2026early}. EgoBrain pairs EEG with egocentric video to evaluate the selective use of EEG when visual information is already available \cite{lin2026egobrain}. Together, these datasets span cognitive monitoring, intention recognition, and visual-neural action understanding under different sensing and fusion settings. Detailed preprocessing, upstream-model settings, and training configurations are provided in Supplementary Section~SI-B. Without a fixed quota, the selected proportion reflected how often VECTA predicted that Fusion would be beneficial. Participant-level BLR and \(\Delta\mathrm{ACC}\) were assessed using two-sided Wilcoxon signed-rank tests with Holm correction across the three datasets. The 95\% confidence intervals were estimated from 50,000 participant-level bootstrap resamples.

Table~\ref{tab:external_evaluation} reports pooled changes relative to Base under the unconstrained rule \(\widehat{G}_{\mathrm{s}} > 0\). VECTA selected 405 of 2,100 RITHM samples (19.3\%). This selection produced a BLR of \(0.00917\), a 3.62\% relative reduction in Brier loss, and a 4.81-pp accuracy gain. On Human-to-Robot Handover, VECTA selected only 18 of 880 samples (2.0\%), with a negligible BLR of \(0.00022\) and no change in accuracy. On EgoBrain, VECTA selected 73 of 137 segments (53.3\%). This selection produced a BLR of \(0.00510\), the largest relative reduction in Brier loss at 11.67\%, and a 2.92-pp accuracy gain.

Fig.~\ref{fig:downstream_performance}(a) shows three distinct coverage-gain patterns. RITHM exhibited low-to-moderate adoption and a positive but heterogeneous mean BLR. Handover clustered near zero adoption and gain. EgoBrain showed substantially broader adoption and greater variation across participants. Fig.~\ref{fig:downstream_performance}(b) shows that the RITHM accuracy gains were distributed across multiple participants and remained significant after correction (\(p_{\mathrm{Holm}} < 0.001\), \(r_{\mathrm{rb}} = 0.852\)). However, its BLR narrowly missed significance (\(p_{\mathrm{Holm}} = 0.0662\)). Neither metric was significant for Handover (\(p_{\mathrm{Holm}} = 0.9219\) for BLR and \(0.7500\) for \(\Delta\mathrm{ACC}\)). For EgoBrain, BLR was significant (\(p_{\mathrm{Holm}} = 0.0309\), \(r_{\mathrm{rb}} = 0.436\)). The mean accuracy increase, however, was concentrated in fewer participants and was nonsignificant after correction (\(p_{\mathrm{Holm}} = 0.2500\)).

Adoption coverage ranged from 2.0\% to 53.3\%. This variation shows that VECTA adapted its use of Fusion to the available sample-level EEG value. The significant accuracy gain on RITHM and the significant reduction in Brier loss on EgoBrain demonstrate that this adaptive selection can provide measurable system-level benefits across distinct multi-modal settings. These contrasting results suggest that VECTA is most useful when the Base modalities leave relevant cognitive or action information unresolved and EEG provides identifiable incremental evidence for a nontrivial subset of samples. When the target may already be well represented by observable behavior, as in the handover task, positive EEG gains become sparse, and VECTA conservatively retains Base for most samples.
\section{Conclusion}
\label{sec:conclusion}

This study investigated when EEG should be used as complementary evidence for an established HRC predictor. We formulated this need as a sample-wise incremental-value estimation problem and proposed VECTA, a decision-aligned framework that selectively adopted EEG-informed Fusion predictions. VECTA used system gain to quantify the decision-level benefit of Fusion over Base and matching gain as an independent audit of the contribution of sample-matched EEG. A structured signed-value learner estimated the otherwise unavailable system gain from prediction states, model disagreement, EEG quality, and temporal contribution. Ranking and policy objectives aligned the estimated value with budget-constrained adoption. Across the three primary datasets, VECTA produced system-gain estimates that significantly tracked realized gains. At a 40\% adoption budget, the selected Fusion outputs captured a disproportionately large share of the full-adoption benefit in all three datasets. VECTA also remained competitive with established non-oracle selectors across different Base strengths and EEG utility profiles. Additional analyses showed that correctly matched EEG carried task-relevant information, although its contribution to the existing system varied across samples. VECTA reduced harmful Fusion adoption, and the ablation results showed that the complete objective improved global gain ordering. In downstream validation using frozen upstream predictors, VECTA yielded measurable gains on RITHM and EgoBrain. It largely retained Base on Human-to-Robot Handover when positive incremental value was scarce.

VECTA was evaluated retrospectively using precomputed upstream predictions, and the adoption decision was made after the Fusion output had been generated. Thus, the current budget regulates the use of Fusion outputs in final decisions but does not account for EEG acquisition or Fusion computation. Future work will evaluate VECTA prospectively in real-time HRC, develop uncertainty-aware value transfer across subjects and tasks, and jointly optimize EEG acquisition, computation, and adoption in multi-class and continuous decision settings.

\bibliographystyle{IEEEtran}
\bibliography{references}

@IEEEtranBSTCTL{BSTcontrol,
  CTLdash_repeated_names = "no"
}

@article{guggenberger2026realtime,
  author      = {Guggenberger, R. and Negahbani, F. and Geb{\"u}hr, J.-S. and Keute, M. and Gharabaghi, A.},
  title       = {Real-time measurement instead of prediction enables high-precision phase-specific stimulation of the human brain},
  journal     = {NeuroImage},
  volume      = {340},
  month       = oct,
  year        = {2026},
  note        = {{Art. no.} 122216, doi: 10.1016/j.neuroimage.2026.122216}
}

@article{elgohr2026trust,
  author      = {Elgohr, A. T. and Rashad, M. and El-Gendy, E. M. and Shaaban, W. and Saafan, M. M.},
  title       = {Trust as a design principle in human--robot collaboration: A review of explainable and adaptive control},
  journal     = {Artif. Intell. Rev.},
  volume      = {59},
  number      = {5},
  year        = {2026},
  note        = {{Art. no.} 134, doi: 10.1007/s10462-026-11554-1}
}

@article{li2023proactive,
  author      = {Li, S. and Zheng, P. and Liu, S. and Wang, Z. and Wang, X. V. and Zheng, L. and Wang, L.},
  title       = {Proactive human--robot collaboration: Mutual-cognitive, predictable, and self-organising perspectives},
  journal     = {Robot. Comput.-Integr. Manuf.},
  volume      = {81},
  month       = jun,
  year        = {2023},
  note        = {{Art. no.} 102510, doi: 10.1016/j.rcim.2022.102510}
}

@article{yao2024task,
  author      = {Yao, B. and Li, X. and Ji, Z. and Xiao, K. and Xu, W.},
  title       = {Task reallocation of human-robot collaborative production workshop based on a dynamic human fatigue model},
  journal     = {Comput. Ind. Eng.},
  volume      = {189},
  month       = mar,
  year        = {2024},
  note        = {{Art. no.} 109855, doi: 10.1016/j.cie.2023.109855}
}

@article{ehrlich2023human,
  author      = {Ehrlich, S. K. and Dean-Leon, E. and Tacca, N. and Armleder, S. and Dimova-Edeleva, V. and Cheng, G.},
  title       = {Human-robot collaborative task planning using anticipatory brain responses},
  journal     = {PLOS ONE},
  volume      = {18},
  number      = {7},
  month       = jul,
  year        = {2023},
  note        = {{Art. no.} e0287958, doi: 10.1371/journal.pone.0287958}
}

@article{ma2025human,
  author      = {Ma, W. and Duan, A. and Lee, H.-Y. and Zheng, P. and Navarro-Alarcon, D.},
  title       = {Human-aware reactive task planning of sequential robotic manipulation tasks},
  journal     = {IEEE Trans. Ind. Informat.},
  volume      = {21},
  number      = {4},
  pages       = {2898--2907},
  month       = apr,
  year        = {2025},
  note        = {doi: 10.1109/TII.2024.3514130}
}

@article{john2024prediction,
  author      = {John, A. R. and Singh, A. K. and Gramann, K. and Liu, D. and Lin, C.-T.},
  title       = {Prediction of cognitive conflict during unexpected robot behavior under different mental workload conditions in a physical human--robot collaboration},
  journal     = {J. Neural Eng.},
  volume      = {21},
  number      = {2},
  month       = mar,
  year        = {2024},
  note        = {{Art. no.} 026010, doi: 10.1088/1741-2552/ad2494}
}

@article{buerkle2021feasibility,
  author      = {Buerkle, A. and Bamber, T. and Lohse, N. and Ferreira, P.},
  title       = {Feasibility of detecting potential emergencies in symbiotic human-robot collaboration with a mobile {EEG}},
  journal     = {Robot. Comput.-Integr. Manuf.},
  volume      = {72},
  month       = dec,
  year        = {2021},
  note        = {{Art. no.} 102179, doi: 10.1016/j.rcim.2021.102179}
}

@article{kalinowska2023embodied,
  author      = {Kalinowska, A. and Pilarski, P. M. and Murphey, T. D.},
  title       = {Embodied communication: How robots and people communicate through physical interaction},
  journal     = {Annu. Rev. Control Robot. Auton. Syst.},
  volume      = {6},
  pages       = {205--232},
  month       = may,
  year        = {2023},
  note        = {doi: 10.1146/annurev-control-070122-102501}
}

@article{wang2024multimodal,
  author      = {Wang, T. and Zheng, P. and Li, S. and Wang, L.},
  title       = {Multimodal human--robot interaction for human-centric smart manufacturing: A survey},
  journal     = {Adv. Intell. Syst.},
  volume      = {6},
  number      = {3},
  month       = mar,
  year        = {2024},
  note        = {{Art. no.} 2300359, doi: 10.1002/aisy.202300359}
}

@article{saleem2025review,
  author      = {Saleem, Z. and Gustafsson, F. and Furey, E. and McAfee, M. and Huq, S.},
  title       = {A review of external sensors for human detection in a human robot collaborative environment},
  journal     = {J. Intell. Manuf.},
  volume      = {36},
  number      = {4},
  pages       = {2255--2279},
  month       = apr,
  year        = {2025},
  note        = {doi: 10.1007/s10845-024-02341-2}
}

@article{angleraud2024sensor,
  author      = {Angleraud, A. and Ekrekli, A. and Samarawickrama, K. and Sharma, G. and Pieters, R.},
  title       = {Sensor-based human--robot collaboration for industrial tasks},
  journal     = {Robot. Comput.-Integr. Manuf.},
  volume      = {86},
  month       = apr,
  year        = {2024},
  note        = {{Art. no.} 102663, doi: 10.1016/j.rcim.2023.102663}
}

@article{jung2024arbitrary,
  author      = {Jung, D. and Bu, S. and Kim, U.},
  title       = {Arbitrary surface contact sensing method for physical human--robot interaction},
  journal     = {IEEE Trans. Ind. Informat.},
  volume      = {20},
  number      = {6},
  pages       = {8274--8283},
  month       = jun,
  year        = {2024},
  note        = {doi: 10.1109/TII.2024.3366972}
}

@article{hoffman2024inferring,
  author      = {Hoffman, G. and Bhattacharjee, T. and Nikolaidis, S.},
  title       = {Inferring human intent and predicting human action in human--robot collaboration},
  journal     = {Annu. Rev. Control Robot. Auton. Syst.},
  volume      = {7},
  pages       = {73--95},
  month       = jun,
  year        = {2024},
  note        = {doi: 10.1146/annurev-control-071223-105834}
}

@article{li2024brain,
  author      = {Li, Y. and Wang, M. and Xie, X. and Chai, W. and Chen, X.},
  title       = {Brain-inspired perception feature and cognition model applied to safety patrol robot},
  journal     = {IEEE Trans. Ind. Informat.},
  volume      = {20},
  number      = {4},
  pages       = {5683--5691},
  month       = apr,
  year        = {2024},
  note        = {doi: 10.1109/TII.2023.3337971}
}

@article{zhang2025mind,
  author      = {Zhang, Y. and Rajabi, N. and Taleb, F. and Matviienko, A. and Ma, Y. and Bj{\"o}rkman, M. and Kragic, D.},
  title       = {Mind meets robots: A review of {EEG}-based brain-robot interaction systems},
  journal     = {Int. J. Hum.-Comput. Interact.},
  volume      = {41},
  number      = {20},
  pages       = {12784--12815},
  year        = {2025},
  note        = {doi: 10.1080/10447318.2025.2464915}
}

@article{wu2023affective,
  author      = {Wu, D. and Lu, B.-L. and Hu, B. and Zeng, Z.},
  title       = {Affective brain--computer interfaces ({aBCIs}): A tutorial},
  journal     = {Proc. IEEE},
  volume      = {111},
  number      = {10},
  pages       = {1314--1332},
  month       = oct,
  year        = {2023},
  note        = {doi: 10.1109/JPROC.2023.3277471}
}

@article{yilmaz2026eeg,
  author      = {Hatipoglu Yilmaz, B. and Yilmaz, C. M. and Kose, C.},
  title       = {{EEG}-based fusion approaches in multimodal emotion recognition: An in-depth review},
  journal     = {Neurocomputing},
  volume      = {666},
  month       = feb,
  year        = {2026},
  note        = {{Art. no.} 132235, doi: 10.1016/j.neucom.2025.132235}
}

@article{hao2026step,
  author      = {Hao, D. and Meng, M. and Gao, Y. and Lou, X. and Kong, W.},
  title       = {Step-wise prompting meets uncertainty-aware dynamic fusion for robust {EEG}-visual emotion recognition},
  journal     = {IEEE Trans. Affect. Comput.},
  volume      = {17},
  number      = {1},
  pages       = {694--707},
  month       = {Jan.--Mar.},
  year        = {2026},
  note        = {doi: 10.1109/TAFFC.2025.3632304}
}

@article{xue2026multimodal,
  author      = {Xue, Q. and Zhang, Y. and Wu, H. and Song, Y.},
  title       = {A large language model-driven framework for multimodal cognitive workload prediction in industrial human--robot collaboration},
  journal     = {Eng. Appl. Artif. Intell.},
  volume      = {181},
  month       = oct,
  year        = {2026},
  note        = {Part 6, {Art. no.} 115694, doi: 10.1016/j.engappai.2026.115694}
}

@article{xu2025neutrl,
  author      = {Xu, C. and Sang, H. and Zhang, C. and Zhou, Y. and Wang, Z. and Li, X. and He, B.},
  title       = {{NeuTRL}: Neural trust-guided reinforcement learning for human-robot collaboration},
  journal     = {IEEE Robot. Autom. Lett.},
  volume      = {10},
  number      = {8},
  pages       = {8155--8162},
  month       = aug,
  year        = {2025},
  note        = {doi: 10.1109/LRA.2025.3583508}
}

@inproceedings{geifman2019selectivenet,
  author      = {Geifman, Y. and El-Yaniv, R.},
  title       = {{SelectiveNet}: A deep neural network with an integrated reject option},
  booktitle   = {Proc. 36th Int. Conf. Mach. Learn. (ICML)},
  series      = {Proc. Mach. Learn. Res.},
  volume      = {97},
  pages       = {2151--2159},
  year        = {2019}
}

@inproceedings{mozannar2020consistent,
  author      = {Mozannar, H. and Sontag, D.},
  title       = {Consistent estimators for learning to defer to an expert},
  booktitle   = {Proc. 37th Int. Conf. Mach. Learn. (ICML)},
  series      = {Proc. Mach. Learn. Res.},
  volume      = {119},
  pages       = {7076--7087},
  year        = {2020}
}

@inproceedings{okati2021differentiable,
  author      = {Okati, N. and De, A. and Gomez-Rodriguez, M.},
  title       = {Differentiable learning under triage},
  booktitle   = {Adv. Neural Inf. Process. Syst. (NeurIPS)},
  volume      = {34},
  pages       = {9140--9151},
  year        = {2021}
}

@inproceedings{narasimhan2022posthoc,
  author      = {Narasimhan, H. and Jitkrittum, W. and Menon, A. K. and Rawat, A. S. and Kumar, S.},
  title       = {Post-hoc estimators for learning to defer to an expert},
  booktitle   = {Adv. Neural Inf. Process. Syst. (NeurIPS)},
  volume      = {35},
  pages       = {29292--29304},
  year        = {2022},
  note        = {doi: 10.52202/068431-2124}
}

@inproceedings{zhang2026coverage,
  author      = {Zhang, Z. and Nguyen, C. C. and Wells, K. and Do, T.-T. and Rosewarne, D. and Carneiro, G.},
  title       = {Coverage-constrained human-{AI} cooperation with multiple experts},
  booktitle   = {Proc. AAAI Conf. Artif. Intell.},
  volume      = {40},
  number      = {21},
  pages       = {18055--18062},
  year        = {2026},
  note        = {doi: 10.1609/aaai.v40i21.38866}
}

@article{tyralis2024review,
  author      = {Tyralis, H. and Papacharalampous, G.},
  title       = {A review of predictive uncertainty estimation with machine learning},
  journal     = {Artif. Intell. Rev.},
  volume      = {57},
  number      = {4},
  month       = apr,
  year        = {2024},
  note        = {{Art. no.} 94, doi: 10.1007/s10462-023-10698-8}
}

@article{gil2026across,
  author      = {Gil, D. and Hern{\'a}ndez-Sabat{\'e}, A.},
  title       = {What is the across-subject performance of your artificial intelligence method for {EEG} analysis?},
  journal     = {Biomed. Signal Process. Control},
  volume      = {120},
  month       = jul,
  year        = {2026},
  note        = {Part A, {Art. no.} 110020, doi: 10.1016/j.bspc.2026.110020}
}

@inproceedings{burges2005learning,
  author      = {Burges, C. J. C. and Shaked, T. and Renshaw, E. and Lazier, A. and Deeds, M. and Hamilton, N. and Hullender, G. N.},
  title       = {Learning to rank using gradient descent},
  booktitle   = {Proc. 22nd Int. Conf. Mach. Learn. (ICML)},
  pages       = {89--96},
  year        = {2005},
  note        = {doi: 10.1145/1102351.1102363}
}

@article{ehrlich2019feasibility,
  author      = {Ehrlich, S. K. and Cheng, G.},
  title       = {A feasibility study for validating robot actions using {EEG}-based error-related potentials},
  journal     = {Int. J. Soc. Robot.},
  volume      = {11},
  number      = {2},
  pages       = {271--283},
  month       = apr,
  year        = {2019},
  note        = {doi: 10.1007/s12369-018-0501-8}
}

@article{ehrlich2018coadaptation,
  author      = {Ehrlich, S. K. and Cheng, G.},
  title       = {Human-agent co-adaptation using error-related potentials},
  journal     = {J. Neural Eng.},
  volume      = {15},
  number      = {6},
  month       = dec,
  year        = {2018},
  note        = {{Art. no.} 066014, doi: 10.1088/1741-2552/aae069}
}

@article{rihet2026robotic,
  author      = {Rihet, M. and Clodic, A. and Sarthou, G. and Sanmarti-Decool, A. and Roy, R. N.},
  title       = {A set of robotic inductive tasks to monitor human cognitive effort},
  journal     = {Int. J. Soc. Robot.},
  volume      = {18},
  month       = jan,
  year        = {2026},
  note        = {{Art. no.} 12, doi: 10.1007/s12369-025-01353-z}
}

@inproceedings{lin2026egobrain,
  author      = {Lin, N. and Wang, Y. and Han, D. and Jiang, W.-B. and Li, J. and Furuta, R. and Sato, Y. and Li, D.},
  title       = {{EgoBrain}: Synergizing minds and eyes for human action understanding},
  booktitle   = {Proc. Int. Conf. Learn. Represent. (ICLR)},
  year        = {2026},
  url         = {https://proceedings.iclr.cc/paper_files/paper/2026/hash/96214e53d0ca519b0236a189bd3b6a18-Abstract-Conference.html}
}

@article{khanna2026early,
  author      = {Khanna, P. and Rajabi, N. and Demir Kanik, S. U. and Kragic, D. and Bj{\"o}rkman, M. and Smith, C.},
  title       = {Early detection of human handover intentions in human--robot collaboration: Comparing {EEG}, gaze, and hand motion},
  journal     = {Robot. Auton. Syst.},
  volume      = {196},
  month       = feb,
  year        = {2026},
  note        = {{Art. no.} 105244, doi: 10.1016/j.robot.2025.105244}
}

@inproceedings{traub2024overcoming,
  author      = {Traub, J. and Bungert, T. J. and L{\"u}th, C. T. and Baumgartner, M. and Maier-Hein, K. H. and Maier-Hein, L. and J{\"a}ger, P. F.},
  title       = {Overcoming common flaws in the evaluation of selective classification systems},
  booktitle   = {Adv. Neural Inf. Process. Syst. (NeurIPS)},
  volume      = {37},
  pages       = {2323--2347},
  year        = {2024},
  note        = {doi: 10.52202/079017-0076}
}

@inproceedings{jitkrittum2023confidence,
  author      = {Jitkrittum, W. and Gupta, N. and Menon, A. K. and Narasimhan, H. and Rawat, A. S. and Kumar, S.},
  title       = {When does confidence-based cascade deferral suffice?},
  booktitle   = {Adv. Neural Inf. Process. Syst. (NeurIPS)},
  volume      = {36},
  pages       = {9891--9906},
  year        = {2023},
  note        = {doi: 10.52202/075280-0431}
}

@article{yu2027dsmt,
  author      = {Yu, X. and Zhou, Y. and Yang, J. and Shi, T. and Cheng, T.},
  title       = {{DS-MTNet}: Structured multi-task {EEG} decoding for human-machine collaboration},
  journal     = {Neurocomputing},
  volume      = {708},
  year        = {2027},
  note        = {{Art. no.} 135222, doi: 10.1016/j.neucom.2026.135222}
}

\end{document}